\documentstyle[aps,epsf]{revtex}

\begin{document}

\title{Causal Bulk Viscous Dissipative Isotropic Cosmologies with Variable
Gravitational and Cosmological Constants}
\author{M. K. Mak\footnote{E-mail:mkmak@vtc.edu.hk}}
\address{Department of Physics, The Hong Kong University of Science and Technology, Clear Water Bay, Hong Kong, P. R. China.}
\author{J. A. Belinch\'on\footnote{E-mail: abelcal@ciccp.es}}
\address{Grupo Inter-Universitario de An\'alisis Dimensional,
Dept. Fisica ETS Arquitectura UPM, Madrid, Spain.}
\author{T. Harko\footnote{E-mail: tcharko@hkusua.hku.hk}}
\address{Department of Physics, The University of Hong Kong,
Pokfulam Road, Hong Kong, P. R. China.}
\maketitle

\begin{abstract}
We consider the evolution of a flat Friedmann-Robertson-Walker Universe, filled
with a causal bulk viscous cosmological fluid, in the presence of variable
gravitational and cosmological constants. The basic equation for the Hubble parameter,
generalizing the evolution equation in the case of constant gravitational coupling and
cosmological term is derived, under the
supplementary assumption that the total energy of the Universe is conserved. By assuming
that the cosmological constant is proportional to the square of the Hubble parameter
and a power law dependence of the bulk viscosity coefficient, temperature and relaxation time
on the energy density of the cosmological fluid, two classes of exact solutions of the field equations are obtained.
In the first class of solutions the Universe ends in an inflationary era, while in the
second class of solutions the expansion of the Universe is noninflationary for all times.
In both models the cosmological "constant" is a decreasing function of time, while the gravitational
"constant" increases in the early period of evolution of the Universe, tending in the
large time limit to a constant value.

\end{abstract}

\section{Introduction}

Recent observations of type Ia supernovae with redshift up to about 
$z\lesssim 1$ provided evidence that we may live in a low mass-density
Universe, with the contribution of the
non-relativistic matter (baryonic plus dark) to the total energy density of the Universe
of the order of $\ \Omega _{m}\sim 0.3$ \cite{Ri98}-\cite{Pe98}. The value of  $\Omega _{m}$ is
significantly less than unity \cite{OsSt95} and consequently either the Universe is
open or there is some additional energy density $\rho $ sufficient to reach
the value $\Omega _{total}=1$, predicted by inflationary theory. 
Observations also show that the deceleration parameter of the Universe $q$
is in the range $-1\leq q<0$, and the present-day Universe undergoes an
accelerated expansionary evolution.

Several physical models have been proposed to give a consistent physical
interpretation to these observational facts. One candidate, and maybe the
most convincing  one for the missing energy is vacuum energy density or cosmological constant $\Lambda $ \cite{We89}.

Since the pioneering work of Dirac \cite{Di38}, who proposed, motivated by the
occurence of large numbers in Universe, a theory with a
time variable gravitational coupling constant $G$, cosmological models with variable $G$
and nonvanishing and variable cosmological term have been intensively investigated
in the physical literature \cite{Ra88}-\cite{Wa93}. In the isotropic cosmological
model of Chen and Wu \cite{ChWu90} it is supposed, in the spirit of quantum cosmology,
that the effective cosmological constant $\Lambda $ varies as $a^{-2}$ (with $a$ the
scale function). In the cosmological model of Lima and Maia \cite{LiMa94} the cosmological constant
$\Lambda =\Lambda \left( H\right) =3\beta H^{2}+3\left( 1-\beta \right)
H^{3}/H_{I}$ is a complicated function of the Hubble parameter $H$, a constant
$\beta $ and an arbitrary time scale $H_{I}^{-1}$, leading to a cosmic history
beginning from an instability of the de Sitter space-time. The cosmological implications
of a time dependence of the cosmological of the form $\Lambda \sim t^{-2}$ have been
considered by Berman \cite{Be91}. Waga \cite{Wa93} investigated flat cosmological models
with the cosmological term varying as $\Lambda =\alpha /a^{2}+\beta H^{2}+\gamma $,
with $\alpha $, $\beta $ and $\gamma $ constants. In this model exact expressions for
observable quantities can be obtained. Nucleosynthesis in decaying-vacuum cosmological
models based on the Chen-Wu ansatz \cite{ChWu90} has been investigated by Abdel-Rahman
\cite{Ra92}. The consistency with the observed helium abundance and baryon asymmetry
allows a maximum vacuum energy close to the radiation energy today. Anisotropic
Bianchi type I cosmological models with variable $G$ and $\Lambda $ have been
analyzed by Beesham \cite{Be93a} and it was shown that in this case there are no
classical inflationary solutions of pure exponential form. Cosmological models
with the gravitational and cosmological constants generalised as coupling scalars
and with $G\sim a^{n}$ have been discussed by Sistero \cite{Si91}. Generalized field
equations with time dependent $G$ and $\Lambda $ have been proposed in \cite{La85} and
\cite{LaPr86} in an attempt to reconcile the large number hypothesis with Einstein's theory of
gravitation. Limits on the variability of $G$ using binary-pulsar data have been
obtained by Damour, Gibbons and Taylor \cite{DaGiTa88}. A detailed analysis of
Friedmann-Robertson-Walker (FRW) Universes in a wide range of scalar-tensor theories of
gravity has been performed by Barrow and Parsons \cite{BaPa97}.
 
Dissipative effects, including both bulk and shear viscosity, are supposed
to play a very important role in the early evolution of the Universe. The
first attempts at creating a theory of relativistic fluids were those of
Eckart \cite{Ec40} and Landau and Lifshitz \cite{LaLi87}. These theories are
now known to be pathological in several respects. Regardless of the choice
of equation of state, all equilibrium states in these theories are unstable
and in addition signals may be propagated through the fluid at velocities
exceeding the speed of light. These problems arise due to the first order
nature of the theory, that is, it considers only first-order deviations from
the equilibrium leading to parabolic differential equations, hence to
infinite speeds of propagation for heat flow and viscosity, in contradiction
with the principle of causality. Conventional theory is thus applicable only
to phenomena which are quasi-stationary, i.e. slowly varying on space and
time scales characterized by mean free path and mean collision time.

A relativistic second-order theory was found by Israel \cite{Is76} and
developed by Israel and Stewart \cite{IsSt76}, Hiscock and Lindblom \cite
{HiLi89} and Hiscock and Salmonson \cite{HiSa91} into what is called
``transient'' or ``extended'' irreversible thermodynamics. In this model
deviations from equilibrium (bulk stress, heat flow and shear stress) are
treated as independent dynamical variables, leading to a total of 14
dynamical fluid variables to be determined. For general reviews on causal
thermodynamics and its role in relativity see \cite{Ma95} and \cite{Ma96}.

Causal bulk viscous thermodynamics has been extensively used for describing
the dynamics and evolution of the early Universe or in an astrophysical
context. But due to the complicated character of the evolution equations,
very few exact cosmological solutions of the gravitational field equations
are known in the framework of the full causal theory. For a homogeneous
Universe filled with a full causal viscous fluid source obeying the relation 
$\xi \sim \rho ^{1/2}$, with $\rho $ the energy density of the cosmological
fluid, exact general solutions of the field equations have been obtained in 
\cite{ChJa97}-\cite{MaHa00b}. It
has also been proposed that causal bulk viscous thermodynamics can model on
a phenomenological level matter creation in the early Universe \cite{MaHa98a},
\cite{MaHa99b}. Exact causal viscous cosmologies with $\xi \sim \rho
^{s},s\neq 1/2$ have been considered in \cite{MaHa99a}.

Because of technical reasons, most investigations of dissipative causal
cosmologies have assumed FRW symmetry (i.e. homogeneity and isotropy) or
small perturbations around it \cite{MaTr97}. The Einstein field equations
for homogeneous models with dissipative fluids can be decoupled and
therefore are reduced to an autonomous system of first order ordinary
differential equations, which can be analyzed qualitatively \cite{CoHo95}, 
\cite{CoHo96}, \cite{PrHeIb01}.

Recent developments in particle physics and cosmology have shown that the
cosmological constant $\Lambda $ ought to be treated as a dynamical quantity
rather than a simple constant. The dynamics of the scale factor in FRW
type models with a variable $\Lambda $ term has been recently revisited for the pefect
fluid case by Overduin and Cooperstock \cite{OvCo98}.

The effects of dissipation as expressed in the form of a non-vanishing bulk viscosity
coefficient in the stress-energy tensor of the matter in cosmological models
with variable $\Lambda $ and $G$ have been considered by several authors \cite{AbVi97}-\cite{ArBe00}.
The role of a transient bulk viscosity in a FRW space-time with decaying vacuum
has been discussed in \cite{AbVi97}. Models with causal bulk viscous cosmological fluid
have been considered recently by Arbab and Beesham \cite{ArBe00}. They obtained both power-law
and inflationary solutions, with the gravitational constant an increasing function of time.

It is the purpose of the present paper to consider the evolution of causal bulk viscous dissipative
cosmological models in the presence of variable gravitational and cosmological constants.
By assuming that $\Lambda $ is a quadratic function of the Hubble parameter $H$ and that
the total energy of the Universe is conserved, the gravitational field equations
can be exactly integrated and two classes of solutions are obtained. The first class
describes a barotropic viscous fluid with arbitrary $\gamma $, while the second solution
corresponds to a stiff cosmological fluid.

The present paper is organised as follows. The basic equations of the model
are written down in Section II. In Section III we present the first class of exact
solutions. The case of the stiff bulk viscous cosmological fluid is considered
in Section IV. In Section V we discuss and conclude our results.

\section{Geometry, field equations and consequences}

We consider that the geometry of the Universe filled with a
bulk viscous cosmological fluid can be described by a spatially flat FRW type metric given by
\begin{equation}\label{1}
ds^{2}=dt^{2}-a^{2}\left( t\right) \left[ dr^{2}+r^{2}\left( d\theta
^{2}+\sin ^{2}\theta d\phi ^{2}\right) \right] . 
\end{equation}

The Einstein gravitational field equations with variable $G$ and $\Lambda $
are:
\begin{equation}
R_{ik}-\frac{1}{2}g_{ik}R=8\pi G(t)T_{ik}+\Lambda (t)g_{ik}.  \label{ein}
\end{equation}

In the following we consider a system of units so that $c=1$.

The bulk viscous effects can be generally described by means of an effective
pressure $\Pi $, formally included in the effective thermodynamic pressure $%
p_{eff}$ \cite{Ma95}. Then in the comoving frame the energy momentum tensor has the
components $T_{0}^{0}=\rho ,T_{1}^{1}=T_{2}^{2}=T_{3}^{3}=-p_{eff}$. For the
line element (\ref{1}) the Einstein field equations give:
\begin{equation}\label{2}
3\left( \frac{\dot{a}}{a}\right) ^{2}=8\pi G(t)\rho +\Lambda (t),  
\end{equation}
\begin{equation}\label{3}
3\frac{\ddot{a}}{a}=-4\pi G(t)\left( 3p_{eff}+\rho \right) +\Lambda (t),
\end{equation}
where a dot denotes the derivative with respect to the time $t$.

Taking the covariant derivative of the Einstein field equations with respect
to the FRW metric (\ref{1}) we obtain the general conservation law in the
presence of variable gravitational and cosmological constants:
\begin{equation}\label{4}
3H\left( p_{eff}+\rho \right) =-\left( \frac{\dot{G}(t)}{G(t)}\rho +\dot{\rho%
}+\frac{\dot{\Lambda}(t)}{8\pi G(t)}\right) , 
\end{equation}
where we have also introduced the Hubble parameter $H=\dot{a}/a$.

Assuming that the total matter content of the Universe is conserved, $%
T_{i;j}^{(mat)j}=0$, the energy density of the matter obeys the usual
conservation law:
\begin{equation}\label{5}
\dot{\rho}+3H\left( p_{eff}+\rho \right) =0.  
\end{equation}

In the presence of the bulk viscous stress $\Pi $, the effective
thermodynamic pressure term becomes $p_{eff}=p+\Pi $, where $p$ is the
thermodynamic pressure of the cosmological fluid. Then the energy
conservation equation (\ref{4}) can be split into two independent equations:
\begin{equation}\label{6}
\dot{\rho}+3H\left( p+\rho \right) =-3\Pi H,  
\end{equation}
\begin{equation}\label{7}
8\pi \dot{G}\rho +\dot{\Lambda}=0.  
\end{equation}

The causal evolution equation for the bulk viscous pressure is given by \cite{Ma95}
\begin{equation}\label{8}
\tau \dot{\Pi}+\Pi =-3\xi H-\frac{1}{2}\tau \Pi \left( 3H+\frac{\dot{\tau}}{%
\tau }-\frac{\dot{\xi}}{\xi }-\frac{\dot{T}}{T}\right) ,  
\end{equation}
where $T$ is the temperature, $\xi $ the bulk viscosity coefficient and $%
\tau $ the relaxation time.

The growth of the total comoving entropy $\Sigma
(t)$ over a proper time interval $(t_{0},t)$ is
\begin{equation}
\Sigma (t)-\Sigma (t_{0})=-\frac{3}{k_{B}}\int_{t_{0}}^{t}\frac{\Pi a^{3}H}{T%
}dt,
\end{equation}
where $k_{B}$ is the Boltzmann constant.

An important observational quantity is the deceleration parameter $q=\frac{d%
}{dt}\left( \frac{1}{H}\right) -1$. The sign of the deceleration parameter
indicates whether the model inflates or not. The positive sign of $q$
corresponds to ``standard'' decelerating models whereas the negative sign
indicates inflation. The deceleration parameter can be expressed as a
function of the thermodynamic, gravitational and  cosmological quantities in the form
\begin{equation}
q=\frac{4\pi G(t)\left( \rho +3p+3\Pi \right) -\Lambda (t)}{8\pi G(t)\rho
+\Lambda (t)}.
\end{equation}

Curvature is described by the tensor field $R_{ijk}^{l}$. It is well-known
that if one uses singular behavior of the components of the tensor or its
derivatives as a criterion for singularites, one gets into trouble since the
singular behavior of components could be due to singular behavior of the
coordinates or tetrad basis rather than the curvature itself. To avoid this
problem, one should examine the scalars formed out of curvature. The
invariants $R_{ij}R^{ij}$ and $R_{ijkl}R^{ijkl}$ (the Kretschmann scalar)
are very useful for the study of the singular behavior of the flat FRW
metric:
\begin{equation}
R_{ij}R^{ij}=12\left( 3H^{4}+3\dot{H}H^{2}+\dot{H}^{2}\right)
\end{equation}
\begin{equation}
R_{ijkl}R^{ijkl}=12\left( 2H^{4}+2\dot{H}H^{2}+\dot{H}^{2}\right)
\end{equation}

In order to close the system of equations (\ref{2}) and (\ref{6})-(\ref{8}) we have
first to give the equation of state for $p$ and specify $T$, $\tau $ and $\xi $. We shall assume the following laws,
\cite{Ma95}:
\begin{equation}\label{9}
p=\left( \gamma -1\right) \rho ,\xi =\alpha \rho ^{s},T=T_{0}\rho ^{\frac{%
\gamma -1}{\gamma }},\tau =\frac{\xi }{\rho }=\alpha \rho ^{s-1},  
\end{equation}
where $1\leq \gamma \leq 2$, $\alpha \geq 0,$ $T_{0}\geq 0$ and $s\geq 0$
are constants. Equations (\ref{9}) are standard in the study of
bulk viscous cosmological models,
whereas the equation $\tau =\xi /\rho $ is a simple procedure to ensure that
the speed of viscous pulses does not exceed the speed of light.

With these choices the general solution of the gravitational field equations
with variable gravitational and cosmological constants still depends on the
functional form of $G$ and $\Lambda $. In the present paper we shall fix the
mathematical form of the cosmological constant assuming that it is a
function of the Hubble parameter only and its time dependence is:
\begin{equation}\label{10}
\Lambda =3\beta H^{2}.  
\end{equation}

This ansatz, initially proposed in \cite{CaLiWa92} on dimensional ground  has
been widely used to study decaying vacuum cosmological models \cite{Wa93}, \cite{ArRa94}-\cite{SaWa93}.

From Eqs. (\ref{2}), (\ref{7}) and (\ref{10}) it follows that in this model
the gravitational constant $G$ and the energy density $\rho $ are given by:
\begin{equation}\label{11}
G=bH^{-n\beta },\rho =\rho _{0}H^{n}, 
\end{equation}
where $\rho _{0}=\frac{3}{4\pi bn}\geq 0$ and $b\geq 0$ are constants and $%
n=2/\left( 1-\beta \right) \geq 0.$ For physically realistic cosmological
models the condition $0<\beta <1$ must hold, in order to assure a time
decreasing energy density of the Universe. One must also assume that $H$ is a
decreasing function of the cosmological time.

With the use of the barotropic equation of state $p=\left( \gamma -1\right)
\rho $ and Eq. (\ref{3}), we obtain
\begin{equation}\label{12}
\dot{H}+\frac{3\gamma }{n}H^{2}+4\pi G\Pi =0.  
\end{equation}

In view of Eq. (\ref{11}), Eq. (\ref{12}) becomes
\begin{equation}\label{13}
\dot{H}+\frac{3\gamma }{n}H^{2}+4\pi bH^{-n\beta }\Pi =0.  
\end{equation}

With the use of Eqs. (\ref{9}) and (\ref{13}), the causal evolution equation
for the bulk viscosity (\ref{8}) leads to the following equation for the Hubble
function $H$:
\begin{equation}\label{14}
\ddot{H}+\left[ 3H+\alpha _{0}H^{n\left( 1-s\right) }\right] \dot{H}+\left( 
\frac{n-4\gamma }{2\gamma }\right) H^{-1}\dot{H}^{2}+\frac{9}{n}\left( \frac{%
\gamma }{2}-1\right) H^{3}+\frac{3\gamma \alpha _{0}}{n}H^{2+n\left(
1-s\right) }=0,  
\end{equation}
where we have denoted $\alpha _{0}=\left( \rho _{0}\right) ^{1-s}/\alpha $.
Eq. (\ref{14}) generalizes the standard evolution equation for causal bulk
viscous models to the case of variable cosmological and gravitational
constants and reduces to it in the case $G=const.$, $\Lambda =const.$.

By introducing the transformations
\begin{equation}\label{15}
H=y^{\frac{2\gamma }{n}},\eta =\int y^{\frac{2\gamma }{n}}dt, 
\end{equation}
Eq. (\ref{14}) reduces to
\begin{equation}\label{16}
\frac{d^{2}y}{d\eta ^{2}}+\left( 3+\alpha _{0}y^{m}\right) \frac{dy}{d\eta }+%
\frac{9}{2\gamma }\left( \frac{\gamma }{2}-1\right) y+\frac{3\alpha _{0}}{2}%
y^{m+1}=0,  
\end{equation}
where we have denoted $m=\frac{2\gamma \left[ n\left( 1-s\right) -1\right] }{%
n}$.

\section{The first class of solutions}

The general solution of the Eq. (\ref{16}) depends on the values of the
parameter $m$, combining the effects of the thermodynamic equation of state
of matter, of bulk viscosity and of the variation of the cosmological
constant. Since a general solution for arbitrary $m$ is difficult to obtain,
we shall limit our study to some particular values of $m$, for which the
general solution Eq. (\ref{16}) can be expressed in an exact analytical form.

As a first class of solutions we consider the solutions generated by the
choice $m=0$, corresponding to $s=\left( 1+\beta \right) /2$, with $0<\beta
<1$ and $1/2<s<1$. In this case the evolution of the bulk viscous pressure
is directly coupled to the time variation of the gravitational constant,
which is described by the parameter $\beta $.

Then the general solution of
Eq. (\ref{16}) can be obtained immediately in the form:
\begin{equation}
y=c_{+}e^{m_{+}\eta }+c_{-}e^{m_{-}\eta },
\end{equation}
where $c_{+}$ and $c_{-}$ are constant of integration and we have denoted $%
m_{\pm }=\frac{-3-\alpha _{0}\pm \sqrt{\alpha _{0}^{2}+18/\gamma }}{2}$.

Therefore we can express the general solution of the gravitational field
equations for a FRW isotropic space-time filled with
a bulk viscous cosmological fluid in the framework of the full
Israel-Stewart-Hiscock causal theory with variable cosmological and
gravitational constants in the following exact parametric form:
\begin{equation}
t-t_{0}=\int \left[ c_{+}e^{m_{+}\eta }+c_{-}e^{m_{-}\eta }\right]
^{-2\gamma /n}d\eta ,
\end{equation}
\begin{equation}
H=\left[ c_{+}e^{m_{+}\eta }+c_{-}e^{m_{-}\eta }\right] ^{2\gamma
/n},a=a_{0}e^{\eta },
\end{equation}
\begin{equation}
\Lambda =3\beta \left( c_{+}e^{m_{+}\eta }+c_{-}e^{m_{-}\eta }\right)
^{4\gamma /n},G=b\left( 3\beta \right) ^{n\beta /2}\Lambda ^{-n\beta
/2},\rho =\rho _{1}\Lambda ^{n/2},
\end{equation}
\begin{equation}
p=\rho _{1}\left( \gamma -1\right) \Lambda ^{n/2},\xi =\alpha \left( \rho
_{1}\right) ^{s}\Lambda ^{sn/2},T=T_{0}\left( \rho _{1}\right) ^{\frac{%
\gamma -1}{\gamma }}\Lambda ^{\frac{n\left( \gamma -1\right) }{2\gamma }%
},\tau =\alpha \left( \rho _{1}\right) ^{s-1}\Lambda ^{\frac{n}{2}\left(
s-1\right) },
\end{equation}
\begin{equation}
q=\gamma \left( \beta -1\right) F\left( \eta \right) -1,
\end{equation}
\begin{equation}
\Pi =-\frac{\left( 1-\beta \right) \gamma }{4\pi b}\left( c_{+}e^{m_{+}\eta
}+c_{-}e^{m_{-}\eta }\right) ^{2\gamma }\left[ \frac{3}{2}+F\left( \eta
\right) \right] ,
\end{equation}
\begin{equation}
\left| \frac{\Pi }{p}\right| =\frac{2}{3}\frac{\gamma }{\gamma -1}\left| 
\frac{3}{2}+F\left( \eta \right) \right|,
\end{equation}
\begin{equation}
\Sigma -\Sigma _{0}=\frac{3\left( 1-\beta \right) \gamma a_{0}^{3}}{%
k_{B}4\pi bT_{0}\left( \rho _{0}\right) ^{\frac{\gamma -1}{\gamma }}}\int
\left( c_{+}e^{m_{+}\eta }+c_{-}e^{m_{-}\eta }\right) ^{2}\left[ \frac{3}{2}%
+F\left( \eta \right) \right] e^{3\eta }d\eta ,
\end{equation}
\begin{equation}
R_{ijkl}R^{ijkl}=24\left( c_{+}e^{m_{+}\eta }+c_{-}e^{m_{-}\eta }\right)
^{8\gamma /n}\left[ 1+\frac{2\gamma }{n}F\left( \eta \right) +\frac{2\gamma
^{2}}{n^{2}}F^{2}\left( \eta \right) \right] ,
\end{equation}
where $t_{0}$, $a_{0}$ and $\Sigma _{0}$ are constants of integration. We
have denoted $F\left( \eta \right) =\frac{c_{+}m_{+}e^{m_{+}\eta
}+c_{-}m_{-}e^{m_{-}\eta }}{c_{+}e^{m_{+}\eta }+c_{-}e^{m_{-}\eta }}$ and $%
\rho _{1}=\frac{\rho _{0}}{\left( 3\beta \right) ^{n/2}}.$

\section{The second class of solutions}

Equation (\ref{16}) has also other solutions describing the dynamics of the
causal bulk viscous FRW universe for appropriate choice of the parameters $m$
and $\gamma $. By means of further substitutions $v=1/u$ and $u=\frac{dy}{%
d\eta }$ the second order differential equation (\ref{16}) can be
transformed into an Abel type differential equation
\begin{equation}\label{17}
\frac{dv}{dy}=\left[ \frac{9}{2\gamma }\left( \frac{\gamma }{2}-1\right) y+%
\frac{3\alpha _{0}}{2}y^{m+1}\right] v^{3}+\left( 3+\alpha _{0}y^{m}\right)
v^{2}.  
\end{equation}

By introducing two new functions $A\left( y\right) =\frac{3+\alpha
_{0}y^{m}/\left( m+1\right) }{9\left( \gamma /2-1\right) /\left( 2\gamma
\right) +3\alpha _{0}y^{m}/2}$ and $B\left( y\right) =\left[ \frac{9\left(
1-\gamma /2\right) y}{2\gamma }-\frac{3\alpha _{0}y^{m+1}}{2}\right] ^{-1}$
allows to rewrite Eq. (\ref{17}) in the general form
\begin{equation}\label{18}
\frac{dv}{dy}=-\frac{v^{3}}{B\left( y\right) }-\left[ \frac{d}{dy}\frac{%
A\left( y\right) }{B\left( y\right) }\right] v^{2}. 
\end{equation}

By introducing a new variable
\begin{equation}\label{19}
\sigma =\frac{1}{v}-\frac{A\left( y\right) }{B\left( y\right) },  
\end{equation}
Eq. (\ref{18}) can be rewritten in the general form
\begin{equation}\label{20}
\frac{dy}{d\sigma }=\sigma B\left( y\right) +A\left( y\right) ,  
\end{equation}
or equivalently
\begin{equation}\label{21}
\frac{dy}{d\sigma }=\frac{\sigma }{\left[ 9\left( 1-\gamma /2\right)
y/\left( 2\gamma \right) -3\alpha _{0}y^{m+1}/2\right] }+\frac{3+\alpha
_{0}y^{m}/\left( m+1\right) }{9\left( \gamma /2-1\right) /\left( 2\gamma
\right) +3\alpha _{0}y^{m}/2}.  
\end{equation}

Since the energy density of the early universe is supposed to be extremely
high, we can consider the extreme limit of a stiff cosmological fluid with
equilibrium pressure equal to the energy density and then the parameter $\gamma
=2$. In the case $m=-2,$ that is $3s=1+\beta $, $0<\beta <1$ and $1/3<s<2/3$%
, Eq. (\ref{21}) becomes a Riccati type differential equation:
\begin{equation}\label{22}
\frac{dy}{d\sigma }=\frac{2}{\alpha _{0}}y^{2}-\frac{2}{3\alpha _{0}}y\sigma
-\frac{2}{3}.
\end{equation}

A particular solution of Eq. (\ref{22}) is given by
\begin{equation}
y=-\sigma \left( \sigma ^{2}/\alpha _{0}+3/2\right) ^{-1}.
\end{equation}

In the following we denote $\Delta \left( \sigma \right) =\left( \sigma
^{2}/\alpha _{0}+3/2\right) ^{-1}$. Therefore the general solution of Eq. (%
\ref{22}) is
\begin{equation}
y=-\sigma \Delta \left( \sigma \right) +\frac{\Delta ^{2}\left( \sigma
\right) e^{-\frac{\sigma ^{2}}{3\alpha _{0}}}}{C_{1}-\frac{2}{\alpha _{0}}\int
\Delta ^{2}\left( \sigma \right) e^{-\frac{\sigma ^{2}}{3\alpha _{0}}%
}d\sigma },
\end{equation}
where $C_{1}$ is an arbitrary constant of integration.

The general solution of the gravitational field equations for a flat FRW
space-time filled with a bulk viscous cosmological fluid and variable $G$
and $\Lambda $ can be obtained in the following exact parametric form, with $%
\sigma $ taken as parameter:
\begin{equation}
t\left( \sigma \right) -t_{0}=-\frac{2}{3\alpha _{0}}\int y^{2\beta
-1}\left( \sigma \right) d\sigma ,
\end{equation}
\begin{equation}
H=y^{2\left( 1-\beta \right) }\left( \sigma \right) ,a=a_{0}\exp \left[ -%
\frac{2}{3\alpha _{0}}\int y\left( \sigma \right) d\sigma \right] ,
\end{equation}
\begin{equation}
\Lambda =3\beta y^{4\left( 1-\beta \right) }\left( \sigma \right)
,G=by^{-4\beta }\left( \sigma \right) ,\rho =\rho _{0}y^{4}\left( \sigma
\right) ,
\end{equation}
\begin{equation}
p=\rho _{0}y^{4}\left( \sigma \right) ,\xi =\alpha \left( \rho _{0}\right)
^{s}y^{4s}\left( \sigma \right) ,T=T_{0}\left( \rho _{0}\right)
^{1/2}y^{2}\left( \sigma \right) ,\tau =\alpha \left( \rho _{0}\right)
^{s-1}y^{4\left( s-1\right) }\left( \sigma \right) ,
\end{equation}
\begin{equation}
q=2\left( 1-\beta \right) \left[ 3-\frac{\sigma }{y\left( \sigma \right) }-%
\frac{\alpha _{0}}{y^{2}\left( \sigma \right) }\right] -1,\Pi =\frac{1-\beta 
}{4\pi b}y^{2}\left( \sigma \right) \left[ 3y^{2}\left( \sigma \right)
-2\sigma y\left( \sigma \right) -2\alpha _{0}\right] ,
\end{equation}
\begin{equation}
\left| \frac{\Pi }{p}\right| =\frac{2}{3}\left| \frac{3y^{2}\left( \sigma
\right) -2\sigma y\left( \sigma \right) -2\alpha _{0}}{y^{2}\left( \sigma
\right) }\right|,
\end{equation}
\begin{equation}
\Sigma -\Sigma _{0}=\frac{\left( 1-\beta \right) a_{0}^{3}}{2\pi
bk_{B}\alpha _{0}T_{0}\left( \rho _{0}\right) ^{1/2}}\int y\left( \sigma
\right) \left[ 3y^{2}\left( \sigma \right) -2\sigma y\left( \sigma \right)
-2\alpha _{0}\right] \exp \left[ -\frac{2}{\alpha _{0}}\int y\left( \sigma
\right) d\sigma \right] d\sigma ,
\end{equation}
\begin{equation}
R_{ijkl}R^{ijkl}=24y^{4-8\beta }\left[ y^{4}+2\left( \beta -1\right)
y^{2}\left( 3y^{2}-y\sigma -\alpha _{0}\right) +2\left( \beta -1\right)
^{2}\left( 3y^{2}-y\sigma -\alpha _{0}\right) ^{2}\right] ,
\end{equation}
where $t_{0}$, $a_{0}$ and $\Sigma _{0}$ are constants of integration.

\section{Discussions and final remarks}

In the present paper we have considered the evolution of a full causal bulk
viscous Universe in the presence of variable gravitational and cosmological
''constants''. We have adopted a particular model for the time variation of
the cosmological constant, assuming that it is a quadratic function of the
Hubble parameter $H$. Consequently, the gravitational constant has also a
power dependence on the Hubble parameter. In both classes of solutions the
evolution of bulk viscosity coefficient is related to that of $G$ and $%
\Lambda $. While the first solution describes a cosmological fluid with
barotropic equation of state for arbitrary $\gamma $, the range of
application of the second solution is restricted to a very high energy
density fluid, obeying the stiff equation of state.

The causal bulk viscous Universe described by the first class of solutions
starts its evolution from a singular state with zero value of the scale
factor $a$, $a\left( t_{0}\right) =0$, and infinite value of the energy
density, $\rho \left( t_{0}\right) \rightarrow \infty $.

The existence of a
singular initial state is also showed by the behavior of the invariant $%
R_{ijkl}R^{ijkl}$, represented in Fig.1, which is singular function of time.

\begin{figure}[h]
\epsfxsize=8cm
\centerline{\epsffile{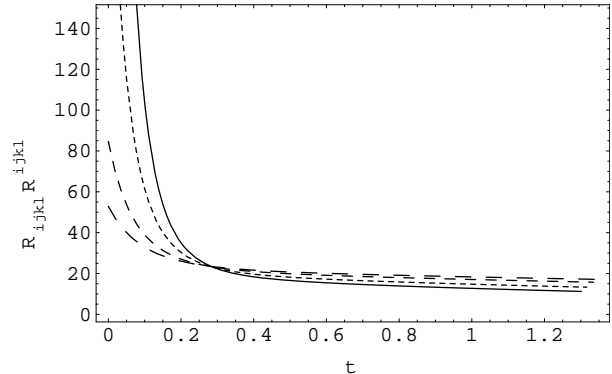}}
\caption{Time variation of the invariant $R_{ijkl}R^{ijkl}$ for the first
class of solutions for a radiation filled bulk viscous Universe ($\gamma =4/3$),
for different values of the parameter $\beta $: $\beta =0.1$ (solid curve), $\beta =0.3$
(dotted curve), $\beta =0.5$ (dashed curve), $\beta =0.6$ (long dashed curve).
We have chosen the integration constants so that $c_{+}=c_{-}=1$ and assumed that
$\alpha _{0}=1$.}
\label{FIG1}
\end{figure}

The evolution of the Universe is expansionary, with the scale factor an
increasing function of time. The Hubble parameter, presented in Fig. 2, is a
monotonically decreasing function of time, similar to the energy density of
the cosmological fluid, represented in Fig. 3.

\begin{figure}[h]
\epsfxsize=8cm
\centerline{\epsffile{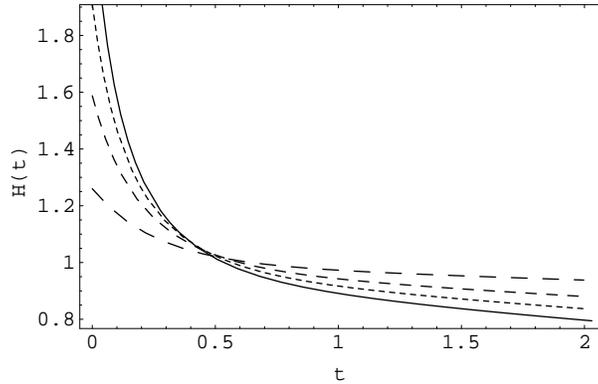}}
\caption{Evolution of the Hubble parameter $H$ for the first
class of solutions for a radiation filled bulk viscous Universe ($\gamma =4/3$),
for different values of the parameter $\beta $: $\beta =0.1$ (solid curve), $\beta =0.3$
(dotted curve), $\beta =0.5$ (dashed curve), $\beta =0.6$ (long dashed curve).
We have chosen the integration constants so that $c_{+}=c_{-}=1$ and assumed that
$\alpha _{0}=1$.}
\label{FIG2}
\end{figure}

\begin{figure}[h]
\epsfxsize=8cm
\centerline{\epsffile{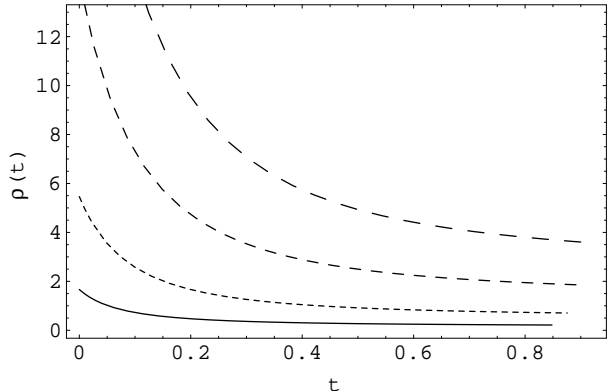}}
\caption{Dynamics of the energy density $\rho $ for the first
class of solutions for a radiation filled bulk viscous Universe ($\gamma =4/3$),
for different values of the parameter $\beta $: $\beta =0.1$ (solid curve), $\beta =0.3$
(dotted curve), $\beta =0.5$ (dashed curve), $\beta =0.6$ (long dashed curve).
We have chosen the integration constants so that $c_{+}=c_{-}=1$ and assumed that
$\alpha _{0}=1$.}
\label{FIG3}
\end{figure}

The bulk viscous pressure $\Pi $, shown in Fig. 4, satisfies the condition $\Pi <0$ only for time
intervals greater than an initial value $t_{1}$.

\begin{figure}[h]
\epsfxsize=8cm
\centerline{\epsffile{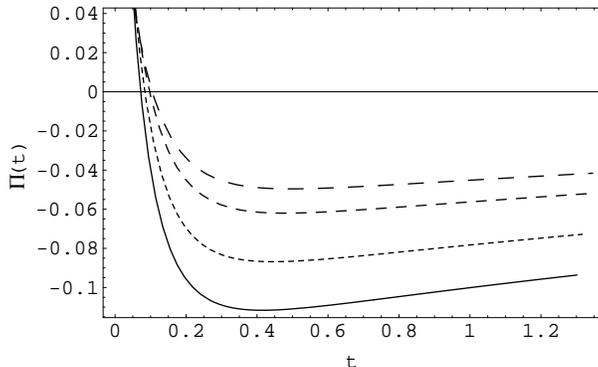}}
\caption{Variation of the bulk viscous pressure $\Pi $ of the cosmological fluid for the first
class of solutions for a radiation filled bulk viscous Universe ($\gamma =4/3$),
for different values of the parameter $\beta $: $\beta =0.1$ (solid curve), $\beta =0.3$
(dotted curve), $\beta =0.5$ (dashed curve), $\beta =0.6$ (long dashed curve).
We have chosen the integration constants so that $c_{+}=c_{-}=1$ and assumed that
$\alpha _{0}=1$.}
\label{FIG4}
\end{figure}

Hence this model can
describe the dynamics of the causal bulk viscous Universe with variable
gravitational and cosmological constants only for a finite time interval.

The deceleration parameter $q$, represented in Fig. 5, shows an initial
non-inflationary evolution, with $q>0$, but in the large time limit the
Universe ends in an inflationary epoch.

\begin{figure}[h]
\epsfxsize=8cm
\centerline{\epsffile{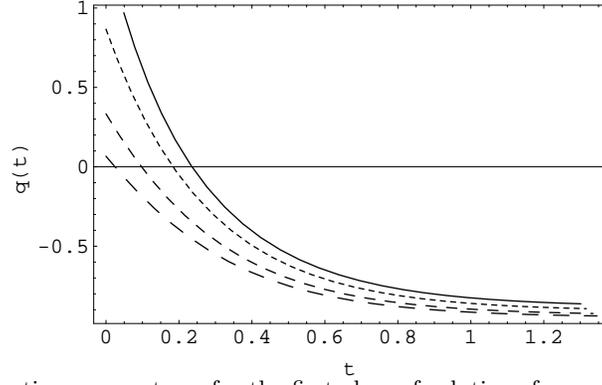}}
\caption{Dynamics of the deceleration parameter $q$ for the first
class of solutions for a radiation filled bulk viscous Universe ($\gamma =4/3$),
for different values of the parameter $\beta $: $\beta =0.1$ (solid curve), $\beta =0.3$
(dotted curve), $\beta =0.5$ (dashed curve), $\beta =0.6$ (long dashed curve).
We have chosen the integration constants so that $c_{+}=c_{-}=1$ and assumed that
$\alpha _{0}=1$.}
\label{FIG5}
\end{figure}

During the viscous effects dominated
era a large amount of comoving entropy is produced. Fig. 6 shows the time
variation of the entropy.

\begin{figure}[h]
\epsfxsize=8cm
\centerline{\epsffile{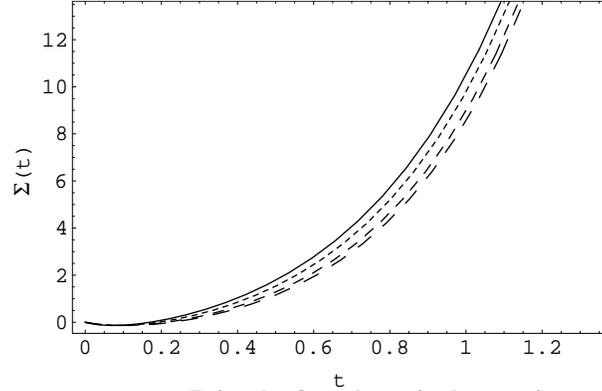}}
\caption{Time variation of the comoving entropy $\Sigma $ for the first
class of solutions for a radiation filled bulk viscous Universe ($\gamma =4/3$),
for different values of the parameter $\beta $: $\beta =0.1$ (solid curve), $\beta =0.3$
(dotted curve), $\beta =0.5$ (dashed curve), $\beta =0.6$ (long dashed curve).
We have chosen the integration constants so that $c_{+}=c_{-}=1$ and assumed that
$\alpha _{0}=1$.}
\label{FIG6}
\end{figure}

The time variation of the cosmological and
gravitational constants is represented in Figs. 7 and 8.

\begin{figure}[h]
\epsfxsize=8cm
\centerline{\epsffile{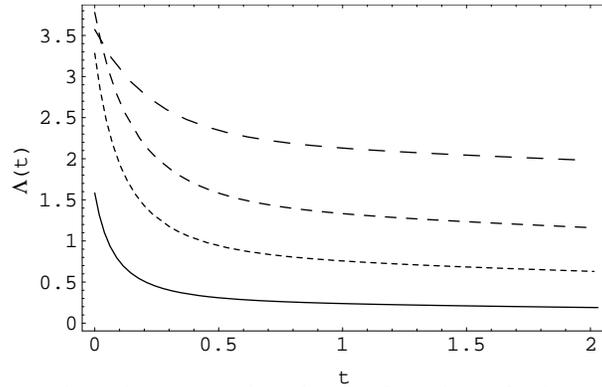}}
\caption{Time variation of the cosmological constant $\Lambda (t)$ for the first
class of solutions for a radiation filled bulk viscous Universe ($\gamma =4/3$),
for different values of the parameter $\beta $: $\beta =0.1$ (solid curve), $\beta =0.3$
(dotted curve), $\beta =0.5$ (dashed curve), $\beta =0.6$ (long dashed curve).
We have chosen the integration constants so that $c_{+}=c_{-}=1$ and assumed that
$\alpha _{0}=1$.}
\label{FIG7}
\end{figure}

\begin{figure}[h]
\epsfxsize=8cm
\centerline{\epsffile{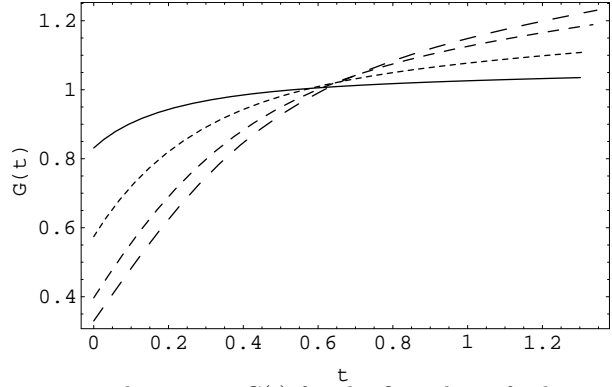}}
\caption{Time variation of the gravitational constant $G(t)$ for the first
class of solutions for a radiation filled bulk viscous Universe ($\gamma =4/3$),
for different values of the parameter $\beta $: $\beta =0.1$ (solid curve), $\beta =0.3$
(dotted curve), $\beta =0.5$ (dashed curve), $\beta =0.6$ (long dashed curve).
We have chosen the integration constants so that $c_{+}=c_{-}=1$ and assumed that
$\alpha _{0}=1$.}
\label{FIG8}
\end{figure}

The cosmological
constant is a decreasing function of time, while the gravitational constant $%
G$ tends in the large time to a constant value. But in this model the ratio
of the bulk viscous pressure and of the thermodynamical presssure is greater
than one during the inflationary period. Consequently during this period the
model is not consistent thermodynamically.

The evolution of the stiff cosmological fluid filled bulk viscous Universe, described by the second class of solutions,
also starts from a singular state, with $a\left( t_{0}\right) =0$. The
invariant $R_{ijkl}R^{ijkl}$, represented in Fig. 9, is also singular in the
large time limit.

\begin{figure}[h]
\epsfxsize=8cm
\centerline{\epsffile{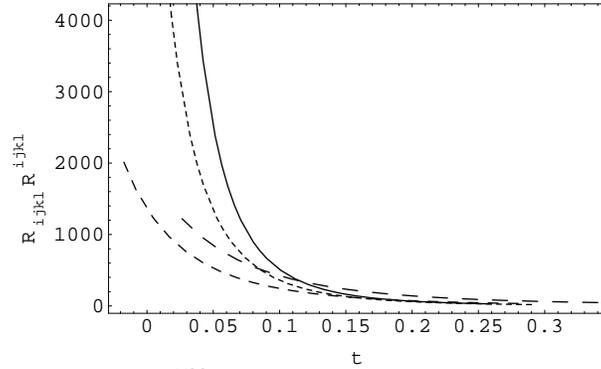}}
\caption{Time variation of the invariant $R_{ijkl}R^{ijkl}$ for the second class
class of solutions describing a stiff fluid filled bulk viscous Universe ($\gamma =2$),
for different values of the parameter $\beta $: $\beta =0.1$ (solid curve), $\beta =0.3$
(dotted curve), $\beta =0.5$ (dashed curve), $\beta =0.6$ (long dashed curve).
We assumed that $\alpha _{0}=1$ and we have chosen the arbitrary integration constant
$C_{1}=1/2$.}
\label{FIG9}
\end{figure}

The evolution of the Universe is expansionary, with the
scale factor, represented in Fig. 10, an increasing function of time.

\begin{figure}[h]
\epsfxsize=8cm
\centerline{\epsffile{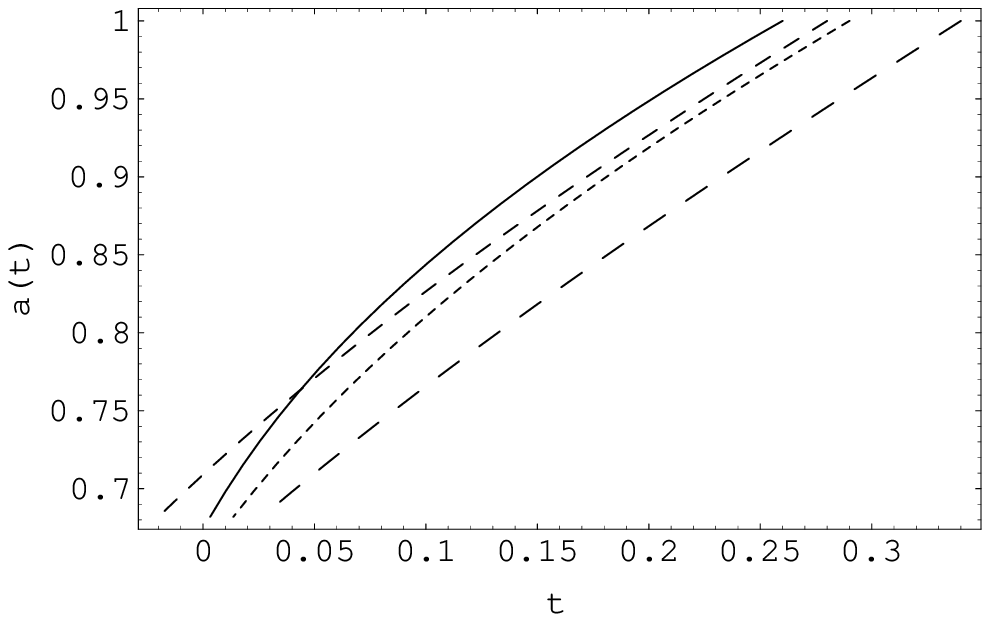}}
\caption{Time evolution of the scale factor $a$ for the second class
class of solutions describing a stiff fluid filled bulk viscous Universe ($\gamma =2$),
for different values of the parameter $\beta $: $\beta =0.1$ (solid curve), $\beta =0.3$
(dotted curve), $\beta =0.5$ (dashed curve), $\beta =0.6$ (long dashed curve).
We assumed that $\alpha _{0}=1$ and we have chosen the arbitrary integration constant
$C_{1}=1/2$.}
\label{FIG10}
\end{figure}

The Hubble parameter and the energy density, represented in Fig. 11, are
monotonically decreasing functions of time.

\begin{figure}[h]
\epsfxsize=8cm
\centerline{\epsffile{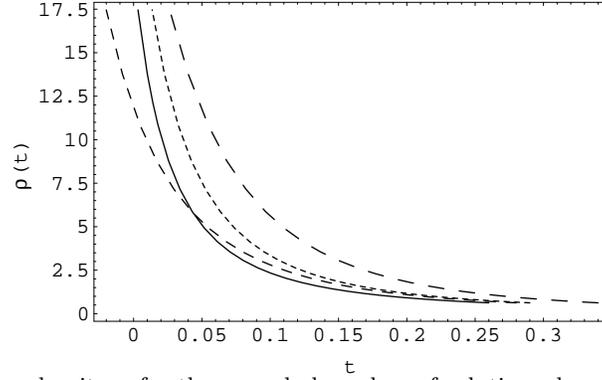}}
\caption{Dynamics of the energy density $\rho $ for the second class
class of solutions describing a stiff fluid filled bulk viscous Universe ($\gamma =2$),
for different values of the parameter $\beta $: $\beta =0.1$ (solid curve), $\beta =0.3$
(dotted curve), $\beta =0.5$ (dashed curve), $\beta =0.6$ (long dashed curve).
We assumed that $\alpha _{0}=1$ and we have chosen the arbitrary integration constant
$C_{1}=1/2$.}
\label{FIG11}
\end{figure}

The bulk viscous pressure $\Pi $
satisfies the condition $\Pi <0$ for all time intervals and for all values
of the parameter, as can be seen from Fig. 12.

\begin{figure}[h]
\epsfxsize=8cm
\centerline{\epsffile{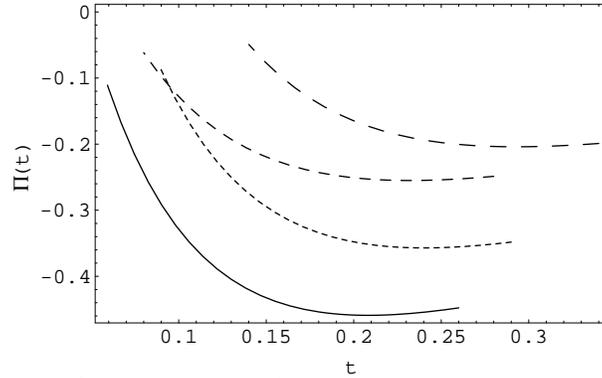}}
\caption{Time variation of the bulk viscous pressure $\Pi $ for the second class
class of solutions describing a stiff fluid filled bulk viscous Universe ($\gamma =2$),
for different values of the parameter $\beta $: $\beta =0.1$ (solid curve), $\beta =0.3$
(dotted curve), $\beta =0.5$ (dashed curve), $\beta =0.6$ (long dashed curve).
We assumed that $\alpha _{0}=1$ and we have chosen the arbitrary integration constant
$C_{1}=1/2$.}
\label{FIG12}
\end{figure}

The deceleration parameter $q$, represented in Fig. 13, satisfies in this model the condition $q>0$ for
all times and in the acceptable range of the physical parameters.

\begin{figure}[h]
\epsfxsize=8cm
\centerline{\epsffile{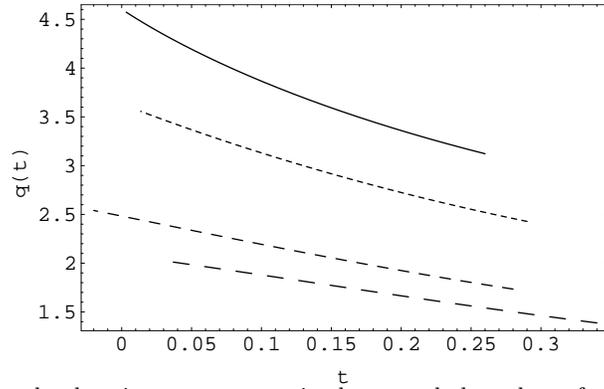}}
\caption{Time dependence of the deceleration parameter $q$ in the second class
class of solutions describing a stiff fluid filled bulk viscous Universe ($\gamma =2$),
for different values of the parameter $\beta $: $\beta =0.1$ (solid curve), $\beta =0.3$
(dotted curve), $\beta =0.5$ (dashed curve), $\beta =0.6$ (long dashed curve).
We assumed that $\alpha _{0}=1$ and we have chosen the arbitrary integration constant
$C_{1}=1/2$.}
\label{FIG13}
\end{figure}

Therefore
the evolution of the high density causal bulk viscous Universe is
non-inflationary. Consequently, the model is consistent thermodynamically
satisfying the condition of the smallness of the bulk viscous pressure. The
time evolutions of the cosmological constant represented in Figs. 14 is similar to
the behavior of this quantity in the first model, with the cosmological
constant a decreasing function of time, while in the large time limit the
gravitational constant, showed in Fig. 15, is a slowly increasing function
of time.

\begin{figure}[h]
\epsfxsize=8cm
\centerline{\epsffile{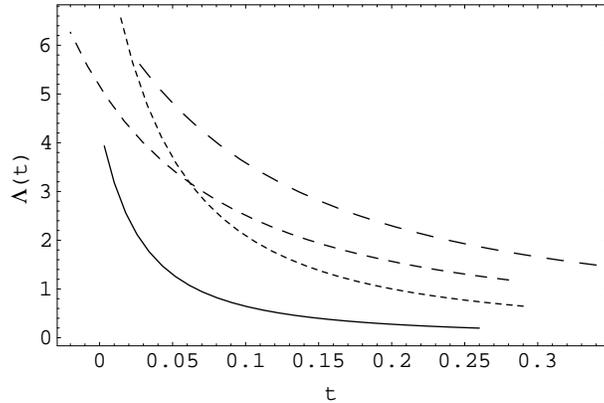}}
\caption{Time variation of the cosmological constant $\Lambda $ for the second class
class of solutions describing a stiff fluid filled bulk viscous Universe ($\gamma =2$),
for different values of the parameter $\beta $: $\beta =0.1$ (solid curve), $\beta =0.3$
(dotted curve), $\beta =0.5$ (dashed curve), $\beta =0.6$ (long dashed curve).
We assumed that $\alpha _{0}=1$ and we have chosen the arbitrary integration constant
$C_{1}=1/2$.}
\label{FIG14}
\end{figure}

\begin{figure}[h]
\epsfxsize=8cm
\centerline{\epsffile{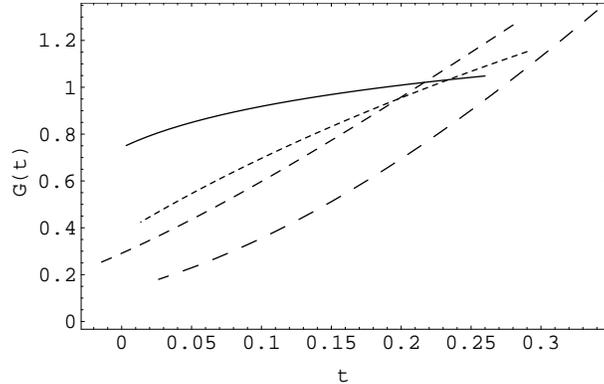}}
\caption{Evolution of the gravitational constant $G(t)$ for the second class
class of solutions describing a stiff fluid filled bulk viscous Universe ($\gamma =2$),
for different values of the parameter $\beta $: $\beta =0.1$ (solid curve), $\beta =0.3$
(dotted curve), $\beta =0.5$ (dashed curve), $\beta =0.6$ (long dashed curve).
We assumed that $\alpha _{0}=1$ and we have chosen the arbitrary integration constant
$C_{1}=1/2$.}
\label{FIG15}
\end{figure}

The possibility that the cosmological constant and the gravitational
coupling are not real constants is an intriguing possibility, which has
intensively been investigated in the physical literature. It is a very
plausible hypothesis that these effects were much stronger in the early
Universe, when dissipative effects also played an important role in the
dynamics of the cosmological fluid. Hence the solutions obtained in the
present paper could give an appropriate description of the early period of
our Universe.

\end{document}